\documentclass[onecolumn]{aastex6}

\usepackage{natbib}
\usepackage{color}
\usepackage{mathtools}
\usepackage{hyperref} %----set up hyper link to citations, equations...
\hypersetup{colorlinks=true,citecolor=blue}

\def	\cm		{\,{\rm {cm}}}

\def	\H		{{\rm H}}

\def	\g		{\,{\rm g}}
\def	\AU		{\,{\rm AU}}
\def	\s		{\,{\rm s}}

\newcommand     \muG    {\,\mu{\rm G}}  % to use in math mode

\def \bea {\begin{eqnarray}}
\def \ena {\end{eqnarray}}               

\begin{document}
\shorttitle{Drag and magnetic forces}
\shortauthors{Hoang and Loeb}

\title{Can Planet Nine Be Detected Gravitationally by a Sub-Relativistic Spacecraft?}

%% Note that the corresponding author command and emails has to come
%% before everything else. Also place all the emails in the \email
%% command instead of using multiple \email calls.
\author{Thiem Hoang}
\affil{Korea Astronomy and Space Science Institute, Daejeon 34055, Republic of Korea; \href{mailto:thiemhoang@kasi.re.kr}{thiemhoang@kasi.re.kr}}
\affil{Korea University of Science and Technology, Daejeon 34113, Republic of Korea}
\author{Abraham Loeb}
\affil{Astronomy Department, Harvard University, 60 Garden Street, Cambridge, MA, USA; \href{mailto:aloeb@cfa.harvard.edu}{aloeb@cfa.harvard.edu}}

%% Note that RNAAS manuscripts DO NOT have abstracts.

%% Start the main body of the article. If no sections in the 
%% research note leave the \section call blank to make the title.

\begin{abstract}
Planet 9 was proposed as an explanation for the clustering of orbits for some trans-Neuptunian objects. Recently, the use of a sub-relativistic spacecraft was proposed to indirectly probe Planet 9's gravitational influence. Here we study the effects of the drag and electromagnetic forces exerted on a sub-relativistic spacecraft by the interstellar medium (ISM) and compare these forces with the gravitational force induced by Planet 9. We find that the resulting noise due to density and magnetic fluctuations would dominate over Planet 9's gravitational signal at sub-relativistic speeds, $v\gtrsim 0.001~c$. We then identify the parameter space required to overcome the drag and magnetic noise from the ISM turbulence and enable the detection of Planet 9's gravity. Finally, we discuss practical strategies to mitigate the effect of the drag and electromagnetic forces.
\end{abstract}

\section{Introduction} 
The clustering of orbits for a group of extreme trans-Neptunian objects (TNOs) suggests the existence of an unseen planet of mass $M\sim 5-10M_{\oplus}$, so-called Planet 9, at a distance of $\sim 400-800$ AU from the Sun (\citealt{Batygin:2019cq}). A primordial black hole was suggested as a substitute for Planet 9 (\citealt{Scholtz:2019wn}). Direct electromagnetic searches have not detected Planet 9 as of yet. 

Recently, \cite{2017ApJ...834L..20C} proposed a novel method to measure the mass of planets via interferometry by an array of relativistic spacecraft, envisioned by the Breakthrough Starshot initiative\footnote{https://breakthroughinitiatives.org/Initiative/3}. \cite{2018AcAau.152..370P} suggested a precursor that will launch slower spacecraft at $v\sim 0.01c$ to explore the Solar system, and \cite{Loeb:2019} mentioned that Planet 9 could be one of its interesting targets. \cite{Witten:2020tc} proposed to use a sub-relativistic spacecraft of speeds $v\gtrsim 0.001c$ to indirectly probe Planet 9 through its gravitational influence on the spacecraft trajectory. \cite{Witten:2020tc} suggested that the small shift of the trajectory along the direction of motion would lead to a detectable time delay. The detection of this effect requires the spacecraft to carry a high-precision clock with an accuracy better than $\lesssim 10^{-5}$ s (equivalent to one part in $10^{12}$ over a period of one year), assuming the spacecraft moves at a speed of $v\gtrsim 0.001$ c. Keeping a high-precision clock on board a lightweight relativistic spacecraft represents a technical challenge for this proposal. To overcome this challenge, \cite{Lawrence:2020ul} considered the transverse effect of gravity and derived the angular deflection of the spacecraft's trajectory to be $\sim 10^{-9}~\rm rad$. They argued that an angular deflection of this magnitude can be measured with an Earth-based or a near-Earth-based telescope and suggested that their method is better than attempting to measure the time delay because the transverse effect is permanent, whereas the time delay is only detectable when the spacecraft passes close to Planet 9. 

Both \cite{Witten:2020tc} and \cite{Lawrence:2020ul} assumed that the spacecraft is moving on a geodesic trajectory from Earth shaped only by gravity, and did not consider the effects of drag or electromagnetic forces  from the interaction of the spacecraft with the interstellar medium (ISM). In this Letter, we compare these effects to the gravitational force induced by Planet 9. In Section \ref{sec:forces}, we describe the drag and magnetic forces and compare them with the gravitational force of Planet 9. In Section \ref{sec:results}, we present our numerical results. In Section \ref{sec:discuss}, we discuss the effects of density and magnetic field fluctuations induced by the ISM turbulence and identify the parameter space required for probing Planet 9 with a sub-relativistic spacecraft. A short summary of our findings is shown in Section \ref{sec:summary}.
%\newpage

\section{Drag and Magnetic Forces from the ISM}\label{sec:forces}
For our present study, we adopt a simple spacecraft design with a cube geometry of width and length $W$ (\citealt{Hoang:2017hg}). The spacecraft mass is $M_{\rm sp}=\rho W^{3}$ where $\rho$ is the mass density, and $A_{\rm sp}=W^{2}=(M_{\rm sp}/\rho)^{2/3}$ is the frontal surface area. For $\rho=3\g\cm^{-3}$, one obtains $M_{\rm sp}\approx 1\g$ for $W\approx 0.7\cm$ and $A_{\rm sp}\approx 0.5\cm^{2}$.

\subsection{Drag Forces}
Since Planet 9 is outside the heliopause of the Solar wind, the spacecraft will encounter the ISM and inevitably experience a drag force due to collisions with gas particles and dust (\citealt{2017ApJ...837....5H}; \citealt{Hoang:2017hg}; \citealt{Lingam:2020io}). For a spacecraft moving at a sub-relativistic speeds $v$ through the ISM with a proton density $n_{\rm H}$ (and 10$\%$ ISM He by abundance), the ratio of the drag force to the gravitational force of Planet 9 equals,
\begin{eqnarray}
\frac{F_{\rm drag}}{F_{\rm grav}}&=&\left(\frac{1.4n_{\rm H}m_{\rm H}v^{2}A_{\rm sp}}{GM_{\rm pl}M_{\rm sp}/b^{2}}\right)=\left(\frac{1.4n_{\rm H}m_{\rm H}v^{2}}{\rho^{2/3}M_{\rm sp}^{1/3}}\right)\left(\frac{b^{2}}{GM_{\rm pl}}\right)\nonumber\\
&\simeq& 117.3\left(\frac{5M_{\oplus}}{M_{\rm pl}}\right)\left(\frac{b}{100 \AU}\right)^{2}\left(\frac{n_{\rm H}}{1\cm^{-3}}\right)\left(\frac{v}{0.01c}\right)^{2}\left(\frac{\rho}{3\g\cm^{-3}}\right)^{-2/3}\left(\frac{M_{\rm sp}}{1\g}\right)^{-1/3},\label{eq:drag_gra}
\end{eqnarray}
where $M_{\rm pl}$ is the mass of the planet, and $b$ is the impact parameter at closest approach to Planet 9. Equation (\ref{eq:drag_gra}) implies dominance of the ISM drag force over gravity for $v\gtrsim 10^{-3}c$ at $b\gtrsim 100\AU$, assuming the typical spacecraft mass of $M_{\rm sp}=1\g$. 

Equation (\ref{eq:def_gra}) implies that the ratio of the forces decreases with increasing spacecraft mass as $M_{\rm sp}^{-1/3}$. The spacecraft mass required for dominance of the gravitational force over the drag force, $F_{\rm grav} > F_{\rm drag}$, is given by,

\bea
M_{\rm sp}> M_{\rm sp,drag}\equiv \left(\frac{1.4n_{\rm H}m_{\rm H}v^{2}b^{2}}{GM_{\rm pl}\rho^{2/3}}\right)^{3}\simeq 1.6\left(\frac{5M_{\oplus}}{M_{\rm pl}}\right)^{3}\left(\frac{b}{100\AU}\right)^{6}\left(\frac{v}{10^{-3}c}\right)^{6}\left(\frac{\rho}{3\g\cm^{-3}}\right)^{-2}\left(\frac{n_{\rm H}}{1\cm^{-3}}\right)^{3}\rm g,\label{eq:Msp}
\ena
where $M_{\rm sp,drag}$ is the critical value for which $F_{\rm grav}=F_{\rm drag}$. Equation (\ref{eq:Msp}) implies a strong dependence of the critical spacecraft mass on its speed $v$ and the impact factor $b$. For a slow mission at $v\sim 10^{-3}c$, the critical mass is $M_{\rm sp,cri}\sim 1.6$ g, but it increases to $M_{\rm sp,drag}\sim 10^{3}$ kg at $v=0.01c$.

\subsection{Electromagnetic Forces}
The spacecraft would inevitably get charged due to collisions with interstellar particles and the photoelectric effect induced by solar and interstellar photons (\citealt{2017ApJ...837....5H}; \citealt{Hoang:2017hg}). The frontal surface layer becomes positively charged through collisions with electrons and protons (i.e., collisional charging), with secondary electron emission being dominant for high-speed collisions (\citealt{Hoang:2017hg}). The outer surface area is charged through the photoelectric effect by ultraviolet photons from the Sun. 

The surface potential, $U$, increases over time due to collisions with the gas and achieves saturation when the potential energy is equal to the maximum energy transfer. One therefore obtains saturation at $eU_{\rm max,e}=m_{e}v^{2}/2$ for impinging electrons and $eU_{\rm max,\rm H}=2m_{e}v^{2}=4eU_{\rm max,e}$ for impinging protons (\citealt{2015ApJ...806..255H}; \citealt{Hoang:2017hg}). The corresponding maximum charge equals
\begin{eqnarray}
Z_{\rm sp,max}\sim \frac{U_{\rm max,\rm H}(W/2)}{e}=\left(\frac{m_{e}v^{2}(M_{\rm sp}/\rho)^{1/3}}{e^{2}}\right)\simeq2.5\times 10^{8}\left(\frac{v}{0.01 c}\right)^{2}\left(\frac{\rho}{3\g\cm^{-3}}\right)^{-1/3} \left(\frac{M_{\rm sp}}{1\rm g}\right)^{1/3},\label{eq:Zmax}
\end{eqnarray}
which implies a strong increase in the maximum charge with spacecraft speed.

The charged spacecraft would therefore experience a Lorentz force due to the interstellar magnetic field, $ F_{B} = eZ_{\rm sp}vB_{\perp}/c$,
where $B_{\perp}$ is the magnetic field component perpendicular to the direction of motion. The ratio of the magnetic force to the gravitational force at an impact parameter $b$ relative to Planet 9 is given by,
\begin{eqnarray}
\frac{F_{\rm mag}}{F_{\rm grav}}&=&\left(\frac{eZ_{\rm sp}vB_{\perp}/c}{GM_{\rm pl}M_{\rm sp}/b^{2}}\right)=\left(\frac{m_{e}v^{3}B_{\perp}}{ec\rho^{1/3}M_{\rm sp}^{2/3}}\right)\left(\frac{b^{2}}{GM_{pl}}\right)\nonumber\\
&\simeq& 6.6\left(\frac{5M_{\oplus}}{M_{\rm pl}}\right)\left(\frac{b}{100 \AU}\right)^{2}\left(\frac{B_{\perp}}{5\muG}\right)\left(\frac{v}{0.01c}\right)^{3}\left(\frac{\rho}{3\g\cm^{-3}}\right)^{-1/3}\left(\frac{M_{\rm sp}}{1\g}\right)^{-2/3},\label{eq:mag_gra}
\end{eqnarray}
where $Z_{\rm sp}=Z_{\rm zp, max}$ is taken, and we adopt the typical magnetic field strength of $B_{\perp}\sim 5~\mu$G as inferred from analysis of the data from Voyager 1 and 2 (\citealt{Opher:2020ib}). The equation above reveals dominance of the magnetic force over the gravitational force for $b\gtrsim 100\AU$ and $v\gtrsim 0.0055c$, assuming the typical spacecraft mass of $M_{\rm sp}=1\g$.

Using the above equation, one obtains the spacecraft mass required for $F_{\rm grav}> F_{\rm mag}$ as follows:
\bea
M_{\rm sp}^{2/3}>M_{\rm sp,mag}^{2/3}\equiv \left(\frac{m_{e}v^{3}B_{\perp}}{ec\rho^{1/3}}\right)\left(\frac{b^{2}}{GM_{\rm pl}}\right)
\ena
which yields
\bea
M_{\rm sp,mag}\simeq 5.7\left(\frac{M_{\rm pl}}{5M_{\oplus}}\right)^{-3/2}\left(\frac{b}{100 \AU}\right)^{3}\left(\frac{B_{\perp}}{5\muG}\right)^{3/2}\left(\frac{v}{0.01c}\right)^{9/2}\left(\frac{\rho}{3\g\cm^{-3}}\right)^{-1/2},
\ena
where $M_{\rm sp,mag}$ is the critical value for which $F_{\rm grav}=F_{\rm mag}$. The above equation implies a strong dependence of the critical mass on the impact parameter and the spacecraft speed.

\subsection{Deflection of Spacecraft's Trajectory by Magnetic Forces}
Next, we estimate the effect of the magnetic deflection on the time delay of the signal. Due to the Lorentz force, the charged spacecraft would move in a curved trajectory instead of a straight line from Earth to Planet 9. The gyroradius of a charged spacecraft,
\begin{eqnarray}
R_{\rm gyro} = \frac{M_{\rm sp}vc}{eZ_{\rm sp}B_{\perp}}\simeq 2.5\times 10^{11}\left(\frac{M_{\rm sp}}{1\g}\right)\left(\frac{v}{0.01c}\right)\left( \frac{10^{9}}{Z_{\rm sp}}\right)\left(\frac{5 \muG}{B_{\perp}}\right)\AU,\label{eq:Rg}
\end{eqnarray}
provides the curvature of the spacecraft trajectory in the magnetic field. 

The deflection in impact parameter from the target at a distance $D\ll R_{\rm gyro}$ is given by,
\begin{eqnarray}
\Delta b= D\alpha \approx \frac{D^{2}}{R_{\rm gyro}}= \frac{D^{2}eZ_{\rm sp}B_{\perp}}{M_{\rm sp}vc}\simeq 10^{-6}\left(\frac{D}{500\AU}\right)^{2} \left(\frac{Z_{\rm sp}}{10^{9}}\right)\left(\frac{1\rm g}{M_{\rm sp}}\right)\left(\frac{B_{\perp}}{5 \muG}\right)\left(\frac{0.01c}{v}\right)\AU,
\label{eq:bmag}
\end{eqnarray}
where $\sin\alpha \approx \alpha\approx D/R_{\rm gyro}$. 

Due to the trajectory deflection, light signals will arrive at a slightly different time than expected without the deflection. The extra distance traversed by the spacecraft is $\Delta x=(\Delta b)\tan \alpha\sim (\Delta b)^{2}/D$ where $\tan \alpha=(\Delta b)/D$. The time delay due to the magnetic deflection is: 
\begin{eqnarray}
\Delta t_{\rm mag}=\frac{\Delta x}{c}=\frac{(\Delta b)^{2}}{Dc}\sim 10^{-12}\left(\frac{\Delta b}{10^{-6}\AU}\right)^{2}\left(\frac{500 \AU}{D}\right)\s.
\end{eqnarray}

The angular deflection of the spacecraft by the interstellar magnetic field is then given by,
\begin{eqnarray}
\alpha_{\rm mag} = \frac{
\Delta b}{D}\simeq 2\times 10^{-9}\left(\frac{\Delta b}{10^{-6}\AU}\right)\left(\frac{500 \AU}{D}\right)\rm rad.
\end{eqnarray}

The time delay due to the gravitational effect was given by \citep{Witten:2020tc},
\begin{eqnarray}
\Delta t_{\rm grav}\simeq 7\times 10^{-7} \left(\frac{M_{\rm pl}}{5M_{\oplus}}\right)\left(\frac{0.01c}{v}\right)^{2}\sinh^{-1}\left(\frac{vt}{b}\right)\s,\label{eq:dt_gra}
\end{eqnarray}
where the origin time $t=0$ is associated with closest approach to Planet 9. The ISM drag force introduces a longitudinal time-delay that exceeds $\Delta t_{\rm grav}$ by the factor given in Equation (\ref{eq:drag_gra}). The gravitational angular deflection was estimated by \cite{Lawrence:2020ul},
\begin{eqnarray}
\alpha_{\rm grav}\simeq 2.8\times 10^{-11}\left(\frac{M_{\rm pl}}{5M_{\oplus}}\right)\left(\frac{100\AU}{b}\right)\left(\frac{0.01c}{v}\right)^{2}\rm rad\ll \alpha_{\rm mag},\label{eq:def_gra}
\end{eqnarray}
which yields a decreasing deflection with spacecraft speed.

\section{Numerical Results}\label{sec:results}

\begin{figure*}
%\centering
\includegraphics[width=0.5\textwidth]{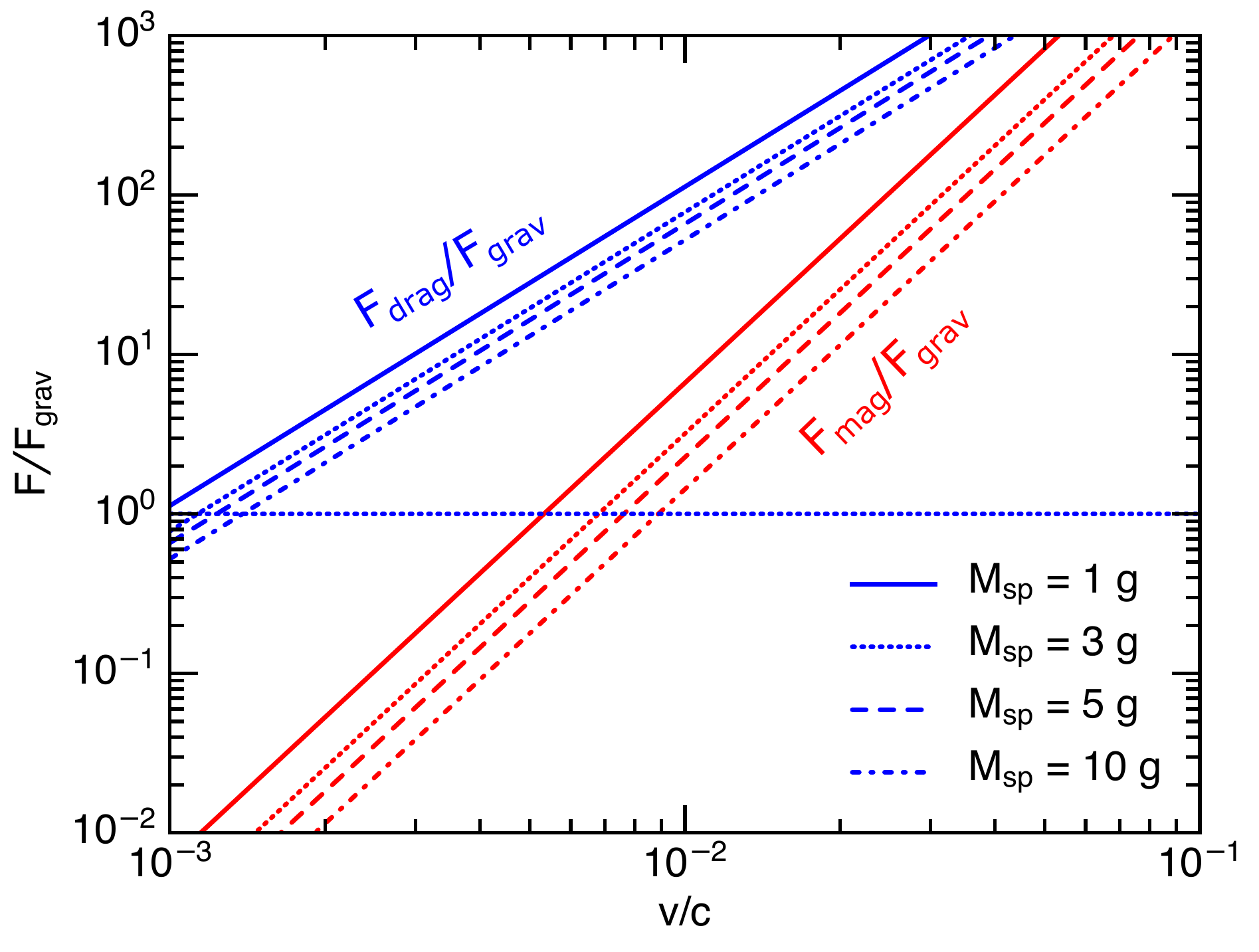}
\includegraphics[width=0.5\textwidth]{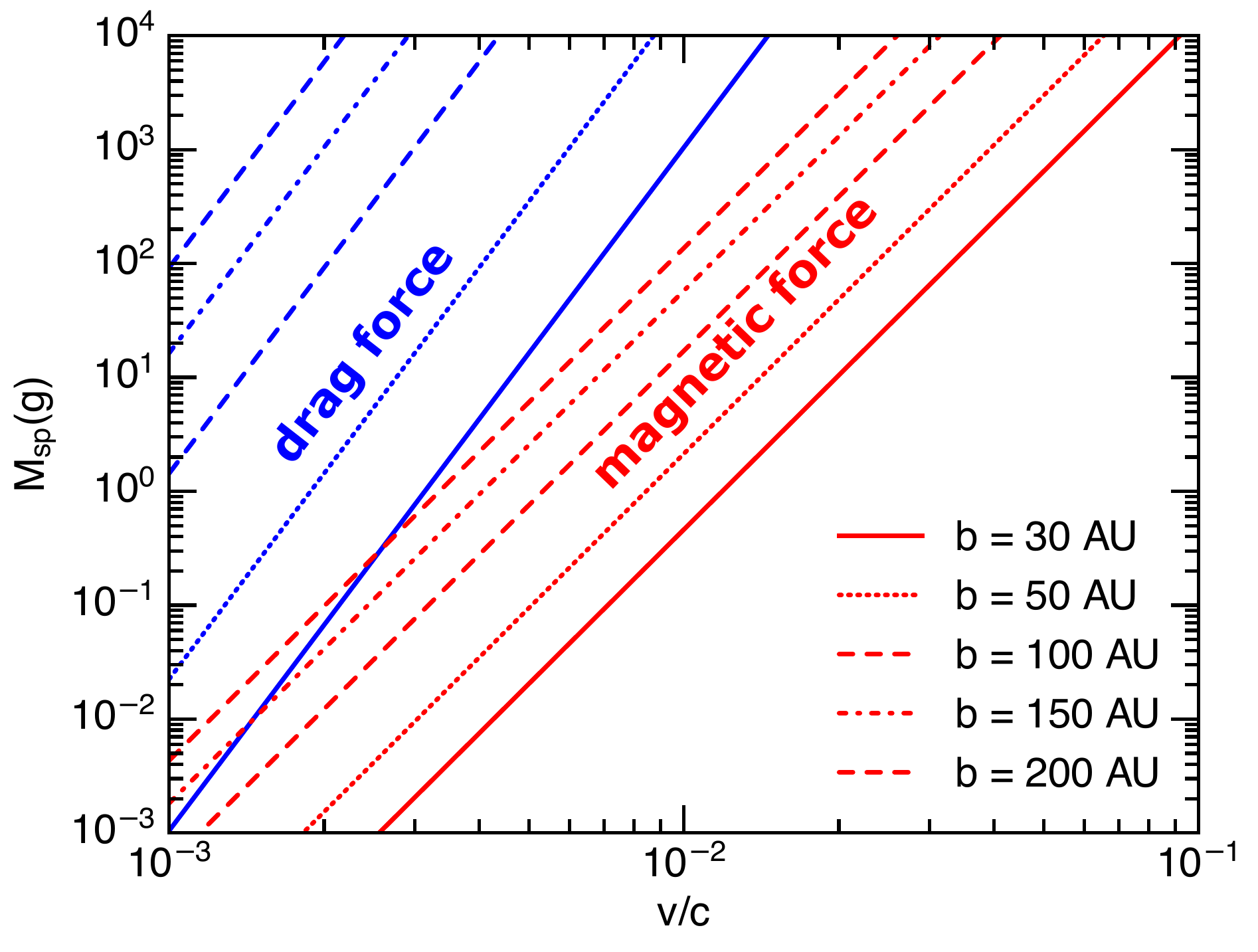}
\caption{Right panel: Ratios of the drag force to the gravitational force of Planet 9 (blue lines) and the magnetic force to the gravitational force (red lines) as a function of the spacecraft speed in units of the speed of light, $v/c$, for different spacecraft masses $M_{\rm sp}$, assuming $M_{\rm pl}=5M_{\oplus}$ and $b=100\AU$. Right panel: Minimum spacecraft mass required for $F_{\rm drag}/F_{\rm grav}=1$ (blue lines) and $F_{\rm mag}/F_{\rm grav}=1$ (red lines) as functions of the spacecraft speed for the different impact parameters, $b$. We assumed $n_{\rm H}=1\cm^{-3}$ and $B_{\perp}=5~\mu$G.}
\label{fig:Forces}
\end{figure*}

In the left panel of Figure \ref{fig:Forces}, we show the ratios between the ISM drag force and Planet 9's gravity (Eq. \ref{eq:drag_gra}) and the magnetic force and gravity (Eq. \ref{eq:mag_gra}) for different spacecraft masses, $M_{\rm sp}$. We assume $M_{\rm pl}=5M_{\oplus}$, $b=100\AU$, and standard parameters for the ISM, with $n_{\rm H}=1\cm^{-3}$ and $B_{\perp}=5~\mu$G (\citealt{Opher:2020ib}). The drag force dominates over gravity for velocities $v\gtrsim 0.001~c$. The magnetic force dominates over gravity only for $v\gtrsim 0.01~c$. Larger spacecraft masses increase the gravitational effect and decrease the force ratios (see also Eq. \ref{eq:drag_gra}). Due to the linear dependence of $F_{\rm drag}$ and $F_{\rm mag}$ on the gas density and magnetic field, density and magnetic fluctuations generic to the ISM turbulence (\citealt{1981Natur.291..561A}) would cause unpredictable fluctuations of the drag and magnetic forces and obscure the signal due to Planet 9's gravity.

In the right panel of Figure \ref{fig:Forces}, we show the minimum spacecraft mass above which $F_{\rm grav}>F_{\rm drag}$ (blue lines) and $F_{\rm grav}>F_{\rm mag}$ (red lines), as functions of the spacecraft speed for the different impact parameters. For a given speed, the spacecraft mass increases rapidly with the impact factor as $b^{6}$. For a small impact parameter of $b=30\AU$, one can send a tiny spacecraft of $M_{\rm sp}\sim 10^{-3}\g$ that can still overcome the drag force. For a large impact parameter of $b=150$ AU, the spacecraft mass must be larger of $M_{\rm sp}\sim 10\g$, assuming $v=0.001c$. For a fleet of spacecraft with a range of impact parameter $b$, the drag force significance would vary from one spacecraft to another.

\section{Discussion}\label{sec:discuss}
We find that due to the interaction with the ISM, a sub-relativistic spacecraft would experience the drag and magnetic forces which will dominate over Planet 9's gravitational influence. In the following, we will discuss in details the effects of density and magnetic fluctuations on the detection of Planet 9's gravitational effect.

\subsection{Effects of Density and Magnetic Field Fluctuations}

%{\bf We can estimate the density fluctuations using scintillation measurements of pulsars (see \cite{Stinebring:2019uq}). 

We quantify the root-mean-square ({\it rms}) fluctuations of the electron density due to the ISM turbulence on a scale $L$ as (\citealt{2011piim.book.....D}, p.115),
\bea
\delta n_{e}\equiv \langle (\Delta n_{e})^{2}\rangle^{1/2}\simeq \left(\frac{6.4\times 10^{-4}}{\cm^{3}}\right)\left(\frac{C_{n}^{2}}{5\times 10^{-17}\cm^{-20/3}}\right)^{1/2}\left(\frac{L}{10^{14}\cm}\right)^{1/3},\label{eq:deltan}
\ena
where $C_{n}^{2}$ is the amplitude of the density power spectrum. Using measurements from Voyager 1 (\citealt{2019NatAs...3..154L}),
$C_{n}^{2}\approx 10^{-2.79}m^{-20/3}\approx 7.52\times 10^{-17}\cm^{-20/3}$ for Equation (\ref{eq:deltan}), yielding $\delta n_{e}\sim 0.0033(L/500\AU)^{1/3}$. With the mean electron density in the local ISM of $n_{e}\sim 0.04\cm^{-3}$ (see e.g., \citealt{2011piim.book.....D}, p.115), one obtains $\delta n_{e}/n_{e}\approx 0.08$.

 %{\bf The properties of local ISM is summarized in \cite{2011AcAau..68..691C}. The Sun is generally accepted to be located in a small local interstellar cloud (LIC) with the mean density of $n_{\H}\sim 0.1-0.2\cm^{-3}$, which is imersed in a local bubble of low density (see \cite{2011AcAau..68..691C}; \citealt{1987ARA&A..25..303C}).}

Assuming $\delta n_{\H}/n_{\H}\sim \delta n_{e}/n_{e}$, one can calculate the ratio of the gravitational signal to noise induced by density fluctuations as follows:
\bea
SNR_{\rm drag} &=& \frac{F_{\rm grav}}{\langle(\Delta F_{\rm drag})^{2}\rangle^{1/2}}=\frac{GM_{\rm pl}M_{\rm sp}/b^{2}}{1.4\delta n_{\H}m_{\H}v^{2}A_{\rm sp}}=\frac{GM_{\rm pl}}{b^{2}}\frac{M_{\rm sp}^{1/3}\rho^{2/3}}{1.4\delta n_{\H}m_{\H}v^{2}}\nonumber\\
&\simeq& 0.085\left(\frac{M_{\rm pl}}{5M_{\oplus}}\right)\left(\frac{b}{100 \AU}\right)^{-2}\left(\frac{M_{\rm sp}}{1\g}\right)^{1/3}\left(\frac{\rho}{3\g\cm^{-3}}\right)^{2/3}\left(\frac{\delta n_{\H}}{0.1n_{\H}}\right)\left(\frac{n_{\rm H}}{1\cm^{-3}}\right)^{-1}\left(\frac{v}{0.01c}\right)^{-2}.\label{eq:SNR_drag}
\ena
Equation (\ref{eq:SNR_drag}) implies the increase of the signal-to-noise ratio with decreasing the impact parameter and spacecraft mass, but the SNR decreases rapidly with increasing spacecraft speed.

To detect the gravitational signal with $SNR_{\rm drag}\gtrsim 3$, the minimum spacecraft mass must satisfy the following condition:
\bea
M_{\rm sp}&\gtrsim &M_{\rm sp,fluc}=\left(3\times \frac{1.4\delta n_{\rm H}m_{\rm H}v^{2}b^{2}}{GM_{\rm pl}\rho^{2/3}}\right)^{3}\left(\frac{SNR_{\rm drag}}{3}\right)^{3}\nonumber\\
&\simeq& 4.8\times 10^{3}\left(\frac{5M_{\oplus}}{M_{\rm pl}}\right)^{3}\left(\frac{b}{100\AU}\right)^{6}\left(\frac{v}{0.01c}\right)^{6}\left(\frac{\rho}{3\g\cm^{-3}}\right)^{-2}\left(\frac{\delta n_{\rm H}}{0.1n_{\H}}\right)^{3}\left(\frac{n_{\rm H}}{1\cm^{-3}}\right)^{3}\left(\frac{SNR_{\rm drag}}{3}\right)^{3}\rm g,
\ena
which indicates a steep increase of the required spacecraft mass with its speed, $v$, and the impact parameter, $b$.

\begin{figure}
\centering
\includegraphics[width=0.6\textwidth]{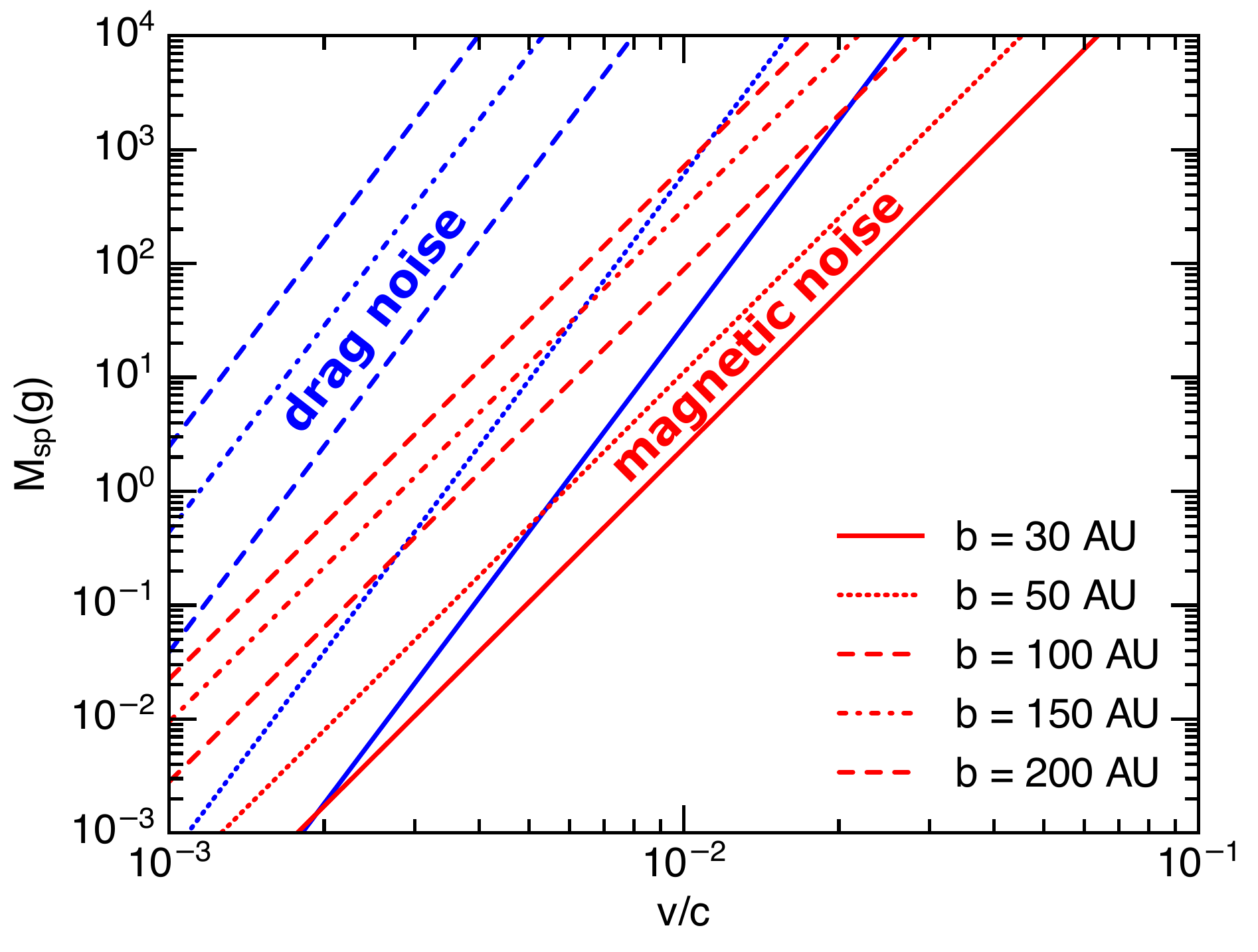}
\caption{The spacecraft mass required for $SNR_{\rm drag,mag}=3$ as a function of the spacecraft speed, $v$, for the different impact parameters, $b$. We assumed $\delta n_{\H}/n_{\H}=0.1$ and $\delta B/B_{\perp}= 1$.}
\label{fig:SNR}
\end{figure}

Figure \ref{fig:SNR} (blue lines) shows the minimum spacecraft mass as a function of the spacecraft speed for the different impact parameters. For a mission of $v=0.001c$ that takes about $\sim 10$ yr to probe Planet 9, the spacecraft mass required to overcome the density fluctuations must increase from $\sim 10^{-3}$ g for $b=50\AU$ to $20$ g for $b=200\AU$. For $v=0.01c$ considered in \cite{2018AcAau.152..370P} that takes $\sim 1$ yr to probe Planet 9, the spacecraft mass must be larger than $10^{3}$ g for $b>50\AU$.

Measuring the angular displacement of the spacecraft, as proposed by \cite{Lawrence:2020ul}, would be particularly challenged by angular deflections from the fluctuating ISM magnetic field. The orientation of the magnetic field is not known in the Planet 9 region of interest. Let $\delta B$ be the {\it rms} magnetic field fluctuations. The ratio of the gravitational signal by Planet 9 to the noise caused by the magnetic field fluctuations is then given by
\bea
SNR_{\rm mag} &=& \frac{F_{\rm grav}}{\langle(\Delta F_{\rm mag})^{2}\rangle^{1/2}}=
\frac{GM_{\rm pl}M_{\rm sp}/b^{2}}{eZ_{\rm sp}v \delta B/c}=
\left(\frac{GM_{\rm pl}M_{\rm sp}^{2/3}}{b^{2}}\right)\left(\frac{ec\rho^{1/3}}{m_{e}v^{3}\delta B}\right)\nonumber
\\
&\simeq&0.15\left(\frac{M_{\rm pl}}{5M_{\oplus}}\right)\left(\frac{b}{100 \AU}\right)^{-2}\left(\frac{M_{\rm sp}}{1\g}\right)^{2/3}\left(\frac{\delta B}{B_{\perp}}\right)^{-1}\left(\frac{B_{\perp}}{5~\mu G}\right)^{-1}\left(\frac{v}{0.01c}\right)^{-3}\left(\frac{\rho}{3\g\cm^{-3}}\right)^{1/3},
\ena
where $Z_{\rm sp}$ was adopted from Equation (\ref{eq:Zmax}), and we assumed $(\delta B)^{2}\sim B^{2}$ as measured by Voyager 1 \citep{2015ApJ...804L..31B}.

To detect the gravitational signal in the presence of magnetic noise with $SNR_{\rm mag}\gtrsim 3$, the minimum spacecraft mass is
\bea
M_{\rm sp}&\gtrsim& M_{\rm sp,fluc}=\left(3\times\frac{m_{e}v^{3}\delta B}{ec\rho^{1/3}}\times\frac{b^{2}}{GM_{\rm pl}}\right)^{3/2}\left(\frac{SNR_{\rm mag}}{3}\right)^{3/2}\nonumber\\
&\simeq&
89.4\left(\frac{M_{\rm pl}}{5M_{\oplus}}\right)^{-3/2}\left(\frac{b}{100 \AU}\right)^{3}\left(\frac{v}{0.01c}\right)^{9/2}\left(\frac{\delta B}{B_{\perp}}\right)^{3/2}\left(\frac{B_{\perp}}{5 \muG}\right)^{3/2}\left(\frac{\rho}{3\g\cm^{-3}}\right)^{-1/2}\left(\frac{SNR_{\rm mag}}{3}\right)^{3/2}~\g.
\ena

Figure \ref{fig:SNR} (red lines) shows $M_{\rm sp, cri}$ as a function of the spacecraft speed for the different impact parameters. For a mission of $v=0.001c$ that takes about $\sim 10$ yr to probe Planet 9, the spacecraft mass required to overcome the magnetic fluctuations must increase from $\sim 0.001$ g for $b=50\AU$ to $0.01$ g for $b=200\AU$. For $v=0.01c$ considered in \cite{2018AcAau.152..370P} that takes $\sim 1$ yr to probe Planet 9, the spacecraft mass must be larger than $1-10^{3}$ g for $b>30-200\AU$.

Note that magnetic field fluctuations are not static, including Alfven waves which are time dependent. This means that the longer the journey is, the larger is the random walk that the spacecraft executes as a result of Alfven waves. This implies a larger noise at spacecraft speeds since they take longer to traverse the vicinity of Planet 9. Finally, during the passage through the heliosphere, the spacecraft would experience large drag and magnetic noise due to the solar wind.

\subsection{Mitigating the Drag and Electromagnetic Forces}
In order to mitigate the effect of the drag force, one can increase the spacecraft mass or decrease the spacecraft speed (see Figure \ref{fig:Forces} and Eq. \ref{eq:drag_gra}). This results in a larger energy cost to launch the spacecraft. A slower spacecraft takes a longer time to reach the Planet 9 region.

One can also design the spacecraft with a needle-like shape to reduce the frontal cross-section. However, as shown in \cite{Hoang:2017hg}, the frontal surface area becomes positively charged and produces an electric dipole. The interaction of the moving dipole with the interstellar magnetic field causes the spacecraft to oscillate around the center of mass, exposing the long axis of the spacecraft to gas collisions and increasing the drag force.

To mitigate the effect of electromagnetic forces, an on-board electron gun could be added. However, this would put additional load on the spacecraft and increases the cost of launch.

%Collisions with dust grains and fluctuations in the spacecraft charge and interstellar magnetic field will all contribute to an unpredictable and time-dependent noise level that is difficult to overcome.}

Since the position of Planet 9 is not known, one could imagine sending a large array of spacecraft so that the impact parameter, $b$, will be minimized for one of them. Unfortunately, due to the fluctuations in the drag force, the spacecraft mass and energy cost must be larger for larger impact parameters $b$. 
%Unfortunately, the spectrum of the ISM turbulence is likely to have a higher amplitude on smaller scales (\citealt{2011piim.book.....D}), making also the noise bigger. 

\section{Summary}\label{sec:summary}
We have studied the drag and electromagnetic forces on the sub-relativistic spacecraft moving in the ISM and compared these forces with the gravitational force produced by Planet 9. We find that the drag force is dominant over gravity for $v\gtrsim 0.001c$ and $b\gtrsim 100\AU$. Density fluctuations on small scales of $\sim 100$ AU represent a critical noise that is difficult to remove for signal retrieval. We identify the critical spacecraft mass for which the gravity is dominant over the drag force. The magnetic force is larger than the gravity for $v\gtrsim 0.005c$ and $b\gtrsim 100\AU$, assuming the typical parameters of the ISM. We identify the spacecraft parameter space required to overcome the drag and magnetic noise from ISM turbulence and to make the detection of Planet 9's gravity possible.

Finally, we noted that recently \cite{2020arXiv200401198Z} claimed that the clustering of orbits of the extreme TNOs may be caused by a dynamical instability and not Planet 9. This proposal requires the outer Solar system to have more mass than previously thought, but there is also the possibility that the clustering is a statistical fluke (\citealt{Clement:2020wk}).

\acknowledgments 
We thank the anonymous referee for helpful comments. We thank Alex Lazarian and Manasvi Lingam for useful comments. This work is supported in part by a grant from the Breakthrough Prize Foundation.

%--------------adding references-----------------------------------
%\bibliographystyle{/Users/thiemhoang/Dropbox/Papers2/apj}
% or other styles: mcbride,plain, abbrv, acm, alpha, apalike, apj
%\bibliography{/Users/thiemhoang/Dropbox/Papers2/cites_paperApJ,/Users/thiemhoang/Dropbox/Papers2/cites_Books}
\bibliography{ms.bbl}

\end{document}